\documentclass[12pt]{article}
\usepackage{epsfig}

\makeatletter
\renewcommand\section{\@startsection {section}{1}{\z@}%
                                   {-3.5ex \@plus -1ex \@minus -.2ex}
                                   {2.3ex \@plus.2ex}%
                                   {\normalfont\large\bfseries}}
\renewcommand\subsection{\@startsection{subsection}{2}{\z@}%
                                     {-3.25ex\@plus -1ex \@minus -.2ex}%
                                     {1.5ex \@plus .2ex}%
                                     {\normalfont\bfseries}}
\makeatother

\def\IZ{\relax\ifmmode\mathchoice
{\hbox{\cmss Z\kern-.4em Z}}{\hbox{\cmss Z\kern-.4em Z}}
{\lower.9pt\hbox{\cmsss Z\kern-.4em Z}} {\lower1.2pt\hbox{\cmsss
Z\kern-.4em Z}}\else{\cmss Z\kern-.4em Z}\fi}
\def\IR{\relax{\rm I\kern-.18em R}}

\def\one{{\hbox{ 1\kern-.8mm l}}}

\newlength{\bredde}
\def\slash#1{\settowidth{\bredde}{$#1$}\ifmmode\,\raisebox{.15ex}{/}
\hspace*{-\bredde} #1\else$\,\raisebox{.15ex}{/}\hspace*{-\bredde}
#1$\fi}

\newcommand  {\Rbar} {{\mbox{\rm$\mbox{I}\!\mbox{R}$}}}

\newsavebox{\zzzbar}
\sbox{\zzzbar}
  {\setlength{\unitlength}{0.9em}
  \begin{picture}(0.6,0.7)
  \thinlines
  \put(0,0){\line(1,0){0.6}}
  \put(0,0.75){\line(1,0){0.575}}
  \multiput(0,0)(0.0125,0.025){30}{\rule{0.3pt}{0.3pt}}
  \multiput(0.2,0)(0.0125,0.025){30}{\rule{0.3pt}{0.3pt}}
  \put(0,0.75){\line(0,-1){0.15}}
  \put(0.015,0.75){\line(0,-1){0.1}}
  \put(0.03,0.75){\line(0,-1){0.075}}
  \put(0.045,0.75){\line(0,-1){0.05}}
  \put(0.05,0.75){\line(0,-1){0.025}}
  \put(0.6,0){\line(0,1){0.15}}
  \put(0.585,0){\line(0,1){0.1}}
  \put(0.57,0){\line(0,1){0.075}}
  \put(0.555,0){\line(0,1){0.05}}
  \put(0.55,0){\line(0,1){0.025}}
  \end{picture}}
\newcommand{\Zbar}{\mathord{\!{\usebox{\zzzbar}}}}

\newcommand{\ena}{\end{eqnarray}}
\newcommand{\beqa}{\begin{eqnarray}}
\newcommand{\eeqa}{\end{eqnarray}}
\newcommand{\bea}{\begin{eqnarray}}
\newcommand{\eea}{\end{eqnarray}}

\newcommand{\eq}[1]{(\ref{#1})}

\newcommand{\be}{\begin{equation}}
\newcommand{\ee}{\end{equation}}

\usepackage{graphicx}

\newcommand{\beq}{\begin{equation}}
\newcommand{\eeq}{\end{equation}}
\newcommand{\ber}{\begin{array}}
\newcommand{\eer}{\end{array}}

\newcommand{\del}{\partial}

\newcommand{\de}{\delta}

\newcommand{\eps}{\varepsilon}

\begin{document}
\begin{titlepage}
\begin{flushright}
arXiv:0706.0824 [hep-th]
\end{flushright}
\vfill
\begin{center}
{\LARGE\bf Quantum evolution across singularities}    \\
\vskip 10mm
{\large Ben Craps and Oleg Evnin}
\vskip 7mm
{\em Theoretische Natuurkunde, Vrije Universiteit Brussel and\\
The International Solvay Institutes\\ Pleinlaan 2, B-1050 Brussels, Belgium}
\vskip 3mm
{\small\noindent  {\tt Ben.Craps@vub.ac.be, eoe@tena4.vub.ac.be}}
\end{center}
\vfill

\begin{center}
{\bf ABSTRACT}\vspace{3mm}
\end{center}

Attempts to consider evolution across space-time singularities often lead to quantum systems with time-dependent Hamiltonians developing an isolated singularity as a function of time. Examples include matrix theory in certain singular time-dependent backgounds and free quantum fields on the two-dimensional compactified Milne universe. Due to the presence of the singularities in the time dependence, the conventional quantum-mechanical evolution is not well-defined for such systems. We propose a natural way, mathematically analogous to renormalization in conventional quantum field theory, to construct unitary quantum evolution across the singularity. We carry out this procedure explicitly for free fields on the compactified Milne universe and compare our results with the
matching conditions considered in earlier work (which were based on the covering Minkowski space).

\vfill

\end{titlepage}
\section{Introduction}

Dynamical evolution across space-time singularities is one of the
most tantalizing, even if speculative, questions in modern theoretical
physics. Should our theories point towards a beginning of time,
it is very natural to ask what came before, and, indeed,
whether there could be anything before.

In certain model contexts, quantum evolution across space-time
singularities appears to be described by time-dependent Hamiltonians
developing an isolated singularity as a function of time at the moment the system
reaches a space-time singularity. It is then worthwhile to study such quantum
Hamiltonians and establish some general prescriptions for using them
to constuct a unitary quantum evolution. Needless to say, additional
specifications are needed in a Schr\"odinger equation involving this
kind of Hamiltonians, on account of the singular time dependence.

One of the simplest examples of such singular time-dependent
Hamiltonians in systems with space-time singularities is given by a
free scalar field on the Milne orbifold (see \cite{Berkooz:2007nm,
Craps:2006yb, Durin:2005ix, Cornalba:2003kd, Tolley:2003nx} and
references therein for some recent occurrences of the Milne orbifold
in models of cosmological singularities). We shall give a detailed
consideration of this case in section~3. Here, it should suffice to
say that the square root determinant of the metric of the Milne
orbifold vanishes as $|t|$ when $t$ goes to 0. Because of that, the
kinetic term in the Lagrangian for a free field $\phi$ on the Milne
orbifold will have the form $|t|(\del_t\phi)^2$, and the
corresponding term in the Hamiltonian expressed through the
canonical momentum $\pi_\phi$ conjugate to $\phi$ will have the form
$\pi_\phi^2/|t|$, which manifestly displays an $1/|t|$ singularity.
The position of this singularity in the time dependence coincides
with the metric singularity of the Milne orbifold.

While it is well-known that free fields on the Milne orbifold are not a good approximation
to interacting systems, especially in gravitational theories \cite{Horowitz:2002mw, Berkooz:2002je},
analogous singular time dependences have recently appeared in other models,
which have been the main motivation for the present work. For example,
11-dimensional quantum gravity with one compact dimension
in a certain singular time-dependent background with a light-like
isometry is conjectured to be described by a time-dependent
modification of matrix string theory \cite{Craps:2005wd,Craps:2006xq}.
This model can be recast in the form of a (1+1)-dimensional
super-Yang-Mills theory on the Milne orbifold. It will thus contain
in its Hamiltonian the $1/|t|$ time dependence typical of the general Milne orbifold
kinematics. The question of transition through the singularity will
then amount to defining a quantum system with such singular Hamiltonian.
Likewise, for the time-dependent matrix models of \cite{Li:2005sz}, which
are conjectured to describe quantum gravity in non-compact eleven-dimensional
time-dependent background with a light-like singularity, one obtains
a quantum Hamiltonian with a singular time dependence.

In view of these examples, our present paper will address the question of
how one should define unitary quantum evolution in the presence of
isolated singularities in the time dependence of quantum
Hamiltonians. Upon giving a general prescription for treating such singularities and discussing the ambiguities it incurs, we shall proceed
with analyzing the simple yet instructive case of a free scalar field
on the Milne orbifold. We shall further discuss the relation between
our prescription and the recipes
for quantum evolution of this system previously proposed in the literature
(and based on considerations in the covering Minkowski space)
\cite{Birrell:1982ix, Nekrasov:2002kf, Tolley:2002cv, Berkooz:2002je}.

\section{Isolated singularities in time-dependent quantum Hamiltonians}

Following the general remarks in the introduction, we shall consider
a quantum system described by the following time-dependent Hamiltonian:
\be
H(t)=f(t,\eps){h}+H_{reg}(t),
\ee
where $H_{reg}(t)$ is non-singular around $t=0$, whereas the numerical function $f(t,\eps)$ develops an isolated singularity at $t=0$ when $\eps$ goes to 0 ($\eps$ serves as a singularity regularization parameter), and $h$ is a time-independent operator. We shall be interested in the
evolution operator from small negative to small positive time. In this region,
we shall assume that we can neglect the regular part of the Hamiltonian $H_{reg}(t)$
compared to the singular part.%
\footnote{
This assumption is actually stronger than one might na\"\i vely have thought: seemingly small
interaction terms in the Hamiltonian are sometimes responsible for large
quantum effects, for instance due to degrees of freedom becoming light. An example where this
happens is the matrix big bang model \cite{Craps:2005wd, Craps:2006xq}, where an important
one-loop potential is generated in the weak coupling region of the field theory. For this reason,
our present discussion will not directly apply to the matrix big bang model, though we hope
to treat that model using similar techniques in future work.
}
The Schr\"odinger equation takes the form
\be
i\frac{d}{dt}|\Psi\rangle=f(t,\eps)h|\Psi\rangle.
\label{schr}
\ee
The solution for the corresponding evolution operator is obviously given by
\be
U(t,t')=\exp\left[-i\int\limits_t^{t'} dt f(t,\eps) h\right].
\label{ev}
\ee
When the regularization parameter $\eps$ is sent to 0, $f(t,\eps)$ becomes singular
and $U(t,t')$ is in general not well-defined.

The goal is then to modify the Hamiltonian locally at $t=0$ in such a way that the evolution away from $t=0$ remains as it was before, but there is a unitary transition
through $t=0$. Of course, a large amount of ambiguity is associated
with such a program, and we shall comment on it below.

The most conservative approach to the Hamiltonian modification is suggested by
(\ref{ev}). Since the problem arises due to the impossibility of integrating
$f(t,\eps)$ over $t$ at $\eps=0$, the natural solution is to modify $f(t,\eps)$
locally around (in the $\eps$-neighborhood of) $t=0$ in such a way that the
integral can be taken (note that we are leaving the operator structure of the
Hamiltonian intact).

The subtractions necessary to appropriately modify $f(t,\eps)$ are familiar
from the theory of distributions. Namely, for any function $f(t,\eps)$
developing a singularity not stronger than $1/t^p$ as $\eps$
is sent to 0, with an appropriate choice of $c_n(\eps)$, one can introduce a modified
\be
\tilde f(t,\eps)=f(t,\eps)-\sum\limits_{n=0}^{p-1}c_n(\eps)\de^{(p)}(t)
\label{distr_subtr}
\ee
(where $\de^{(p)}(t)$ are derivatives of the $\de$-function) in such a way
that the $\eps\to 0$ limit of $\tilde f(t,\eps)$ is defined in the sense
of distributions. The latter assertion would imply that the $\eps\to 0$ limit of
\be\int \tilde f(t,\eps) {\cal F}(t) dt\ee
is defined for any smooth ``test-function'' ${\cal F}(t)$, and, in particular, that the $\eps\to 0$ limit of (\ref{ev}) becomes well-defined,
if $f(t,\eps)$ is replaced by $\tilde f(t,\eps)$. (Note that, since $f(t,\eps)$ and $\tilde f(t,\eps)$ only differ in an infinitesimal neighborhood of $t=0$, this modification will
not affect the evolution at finite $t$).

As a matter of fact, the subtraction needed for our particular case is simpler than
(\ref{distr_subtr}). Since the $n>0$ terms in (\ref{distr_subtr}) can only affect
the value of the evolution operator (\ref{ev}) at $t'=0$, if one is only interested
in the values of the wave function for non-zero times, one can simply omit the $n>0$ terms from (\ref{distr_subtr}). One can then write down the subtraction explicitly as
\be
\tilde f(t,\eps)=f(t,\eps)-\left(\int\limits_{-t_0}^{t_0} f(t,\eps) dt\right)\de(t).
\ee
The appearance of a free numerical parameter (which can be chosen as $t_0$ in the expression above, or a function thereof) is not surprising, since, if $\tilde f(t,\eps)$ is an adequate modification of $f(t,\eps)$, so is $\tilde f(t,\eps) + c\de(t)$ with any finite $c$.

For the particular $1/|t|$ time dependence of the Hamiltonian
mentioned in the introduction, one can choose $f(t,\eps)$ as
$1/\sqrt{t^2+\eps^2}$, in which case $\tilde f(t,\eps)$ becomes
\be\label{fabst}
f_{1/|t|}(t,\eps)={1\over\sqrt{t^2+\eps^2}}+2\ln(\mu\eps)\de(t). \ee
It is sometimes more appealing to replace the $\delta$-function in
\eq{fabst} by a resolved $\delta$-function, in which case we find
\be
f_{1/|t|}(t,\eps)={1\over\sqrt{t^2+\eps^2}}+2\ln(\mu\eps){\eps\over\pi(t^2+\eps^2)}
\label{f1/t} \ee (with $\mu$ being an arbitrary mass scale).

One should note that it is very natural to think of the above subtraction procedure
as renormalizing the singular time dependence of the Hamiltonian. Indeed, the mathematical
structure behind generating distributions by means of $\de$-function subtractions is
precisely the same as the one associated with subtracting local counter-terms
in order to render conventional field theories finite. For concreteness, consider the one-loop contribution to the full momentum space propagator in $\lambda\phi^3$ field theory, given by the diagram
\begin{center}
\begin{picture}(150,42)
\put(0,0){\epsfig{file=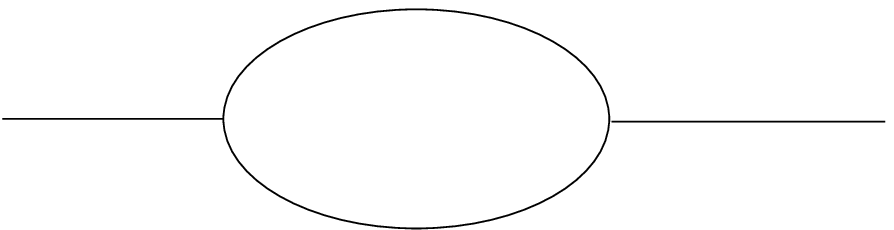,width=5cm}}
\put(28,8){$x$}
\put(105,7){$x'$}
\end{picture}
\end{center}
If we compute it using position space Feynman rules, we find that it is proportional to the
Fourier transform of the square of the scalar field Feynman propagator $D(x,x')$. However, while the Feynman propagator
itself if a distribution, its square is not. For that reason, if one tries to evaluate the Fourier transform, one obtains infinities, since integrals of $[D(x,x')]^2$ cannot be evaluated.
The problem is resolved by subtracting local counter-terms from the field theory
Lagrangian, which, for the above diagram, would translate into adding $\de(x-x')$
and its derivatives (with divergent cutoff-dependent coefficients) to $[D(x,x')]^2$ in such a way
as to make it a distribution. The mathematical structure of this procedure is
precisely the same as what we employed for renormalizing the singular time
dependences in time-dependent Hamiltonians.

We should remark upon the general status of our Hamiltonian prescription
viewed against the background of all possible singularity transition recipes one could devise.
If the only restriction is that the evolution away from the singularity is
given by the original Hamiltonian, one is left with a tremendous infinitefold ambiguity:
any unitary transformation can be inserted at $t=0$ and the predictive power is lost
completely. One should look for additional principles in order to be able to define
a meaningful notion of singularity transition.

Our prescription can be viewed as a very conservative approach,
since it preserves the operator structure of the Hamiltonian (the
counter-terms added are themselves proportional to $h$, the singular
part of the Hamiltonian). In the absense of further physical
specification, this approach appears to be natural and can be viewed
as a sort of ``minimal subtraction''. However, under some
circumstances, one may be willing to pursue a broader range of
possibilities for defining the singularity transition. For example,
one may demand that the resolution of the singular dynamics must
have a geometrical interpretation (at finite values of $\eps$). This
question will be addressed in \cite{CDE}.

In section 3, the focus of our attention will be a particular
quantum system with a Hamiltonian quadratic in the canonical
variables. For such linear systems, it is most common to analyze
quantum dynamics in the Heisenberg picture, rather than in the
Schr\"odinger picture we have employed above for the purpose of describing our
general formalism. For convenience, we shall give a summary
of the relevant derivations in appendix A. In short, one should construct
the most
general {\em classical} solution of the system in the form
\be
x(t)=Au(t)+A^*u^*(t).
\ee
The solution to the Heisenberg equations of motion is simply
obtained by replacing the integration constants $A$ and $A^*$
in the above expression by creation-annihilation operators $a$ and $a^\dagger$,
which (with an appropriate normalization of $u(t)$) satisfy
the standard commutation relation $[a,a^\dagger]=1$. The question
of solving for the quantum dynamics is then most commonly phrased
in terms of constructing the mode functions $u(t)$ and $u^*(t)$,
which are normalized solutions to the classical equations of motion.

Our prescription may equally well be applied in such setting.
One can analyze the classical equations of motion derived from
the time-dependent Hamiltonian. It is safest to do so at finite $\eps$,
since the na\"\i ve $\eps\to0$ limit of the classical equations
of motion may not necessarily exist. However, the $\eps\to 0$ limit
of the solutions for the mode functions {\em will} exist, and will,
of course, define the same quantum dynamics as the general solution
to the Schr\"odinger equation given by (\ref{schr}).

\section{Free fields on the compactified Milne universe}

\subsection{The compactified two-dimensional Milne universe}

The two-dimensional Milne universe
\be\label{MilneMetric}
ds^2=-dt^2+t^2dx^2,
\ee
with $0<t<+\infty$, corresponds to the ``future'' quadrant $X^\pm>0$ of Minkowski space $ds^2=-2dX^+dX^-$ via the identification
\be
X^\pm={1\over\sqrt2}te^{\pm x}.
\ee
The Milne universe can be compactified by the identification
\be\label{compac}
x\sim x+2\pi,
\ee
which corresponds to the discrete boost identification
\be\label{discreteboost}
X^\pm\sim e^{\pm 2\pi}X^\pm.
\ee
The resulting space is a cone, which is singular at its tip $t=0$.

The action for a free scalar field in the (compactified) Milne
universe is \be\label{actionscalar} S=\int
dt\,dx\,t\,\left({\dot\phi^2\over 2}-{{\phi^\prime}^2\over
2t^2}-{m^2\phi^2\over2}\right). \ee The corresponding equation of
motion \be\label{eomMilne} \ddot\phi+{\dot\phi\over t}-{\phi''\over
t^2}+m^2\phi=0 \ee is solved by \bea
\psi_{m,l}(t,x)&=&{1\over2\sqrt2\pi i}\int_{\Rbar}dw\, e^{i\left({m\over\sqrt2}X^-e^{-w}+{m\over\sqrt2}X^+e^w+lw\right)}\label{Nekr}\\
&=&{1\over2\sqrt2}e^{l\pi\over2}e^{-ilx}H^{(1)}_{-il}(mt)
\eea
and their complex conjugates \cite{Birrell:1982ix, Nekrasov:2002kf}. Here $H^{(1)}$ denotes a Hankel function, and the compactification \eq{compac} enforces the momentum quantization condition $l\in\Zbar$.
For solutions to the equation of motion \eq{eomMilne}, we define the scalar product \cite{Birrell:1982ix}
\be\label{scalarproduct}
(\phi_1,\phi_2)=-i\int_0^{2\pi}dx\,t\,\left[\phi_1(t,x)\dot\phi_2^*(t,x)-\dot\phi_1(t,x)\phi_2^*(t,x)\right]
\ee
and the Klein-Gordon norm $(\phi,\phi)$. The solutions \eq{Nekr} are normalized to have Klein-Gordon norm $-1$.

To quantize the scalar field $\phi$, one expands
\be\label{expansion}
\phi(t,x)=\sum_{k\in\Zbar}\left[a_k u_k(t,x)+a_k^\dagger u_k^*(t,x)\right],
\ee
where the $u_k(x,t)$ have Klein-Gordon norm $1$, which ensures the canonical commutation relations
\be
[a_k,a_l^\dagger]=\delta_{k,l}.
\ee
We choose
\be\label{modefunctions}
u_k(t,x)=\psi^*_{m,k}(t,x).
\ee
Essentially because $\psi_{m,k}$ of (\ref{Nekr}) are superpositions of negative
frequency waves on the covering Minkowski space, the vacuum state defined with
the creation and annihilation operators of (\ref{expansion}) is an adiabatic vacuum of infinite order \cite{Birrell:1982ix}. Note, however, that in
a compactified Milne universe (where globally defined inertial frames
are absent) this particular adiabatic vacuum is no more special than any other
adiabatic vacuum of infinite order (of which there are infinitely many).

Near $t=0$, the $l\neq0$ mode functions behave as (see, for instance, \cite{Berkooz:2002je})
\be\label{hnkl}
u_{l}\sim{e^{ilx}\over 2\sqrt{2\pi l\sinh(\pi l)}}\left[-\left( {mt\over2}  \right)^{il}e^{-{\pi l\over2}-i\varphi_l}+ \left( {mt\over2}  \right)^{-il}e^{{\pi l\over2}+i\varphi_l}   \right],
\ee
with $\varphi_l$ defined by $e^{i\varphi_l}=\Gamma(1+il)\sqrt{\frac{\sinh(\pi l)}{\pi l}}$ and satisfying $\varphi_{-l}=-\varphi_l$, while
\be
u_{0}\sim{1\over2\sqrt2}\left(1-{2i\over\pi}\log\left({mt\over2}\right)\right).
\ee
The mode functions are clearly singular at $t=0$. The question we now want to address is whether quantum mechanical evolution can be consistently and naturally defined beyond $t=0$.

In the literature (see, for instance, \cite{Nekrasov:2002kf, Tolley:2002cv, Berkooz:2002je}), this question has been addressed by extending the range of the $t$ coordinate in the compactified Milne metric \eq{MilneMetric} to $-\infty<t<\infty$, i.e.\ by adding a ``past cone'' to the ``future cone''.%
\footnote{
In some string theory contexts, it is natural to consider the full Minkowski space up to the discrete boost identification \eq{discreteboost}, which in addition adds ``whisker'' regions with closed timelike curves. We will not consider whisker regions in the present paper.
}
In the action \eq{actionscalar}, the factor $t$ is replaced by $|t|$,
\be
\label{actionscalarbis}
S=\int dt\,dx\,|t|\,\left({\dot\phi^2\over 2}-{{\phi^\prime}^2\over 2t^2}-{m^2\phi^2\over2}\right).
\ee
The same goes for the scalar product \eq{scalarproduct}
\be\label{scalarproductbis}
(\phi_1,\phi_2)=-i\int_0^{2\pi}dx\,|t|\,\left[\phi_1(t,x)\dot\phi_2^*(t,x)-\dot\phi_1(t,x)\phi_2^*(t,x)\right]
\ee
and the corresponding Klein-Gordon norm.

The question then is how to define matching conditions between $t<0$ mode functions and $t>0$ mode functions, i.e.\ how to define global mode functions. Natural globally defined mode functions are obtained by allowing $X^\pm$ to be either both positive or both negative in \eq{Nekr} (see \eq{modefunctions}). As these are superpositions of negative frequency Minkowski modes, they describe excitations above the (adiabatic) vacuum inherited from Minkowski space. The solutions \eq{Nekr} have the property that they are analytic in the lower complexified $t$-plane. For $t<0$, they can be written as
\be
\psi_{m,l}(t,x)=-{1\over2\sqrt2}e^{-{l\pi\over2}}e^{-ilx}\left(H^{(1)}_{il}(|mt|)\right)^*,\ \ \ \ \ \ \ \ \ \ \ (t<0)
\ee
which still has Klein-Gordon norm $-1$.
For $t$ approaching 0 from below, we have for the corresponding mode functions $u_l(t,x)=\psi^*_{m,l}(t,x)$ with $l\neq0$,
\be\label{hankel}
u_{l}\sim{e^{ilx}\over 2\sqrt{2\pi l\sinh(\pi l)}}\left(-\left| {mt\over2}\right|^{il}e^{{\pi l\over2}-i\varphi_l}+ \left| {mt\over2}  \right|^{-il}e^{-{\pi l\over2}+i\varphi_l}   \right),
\ee
and
\be
u_{0}\sim-{1\over2\sqrt2}\left(1+{2i\over\pi}\log\left|{mt\over2}\right|\right).
\ee

Note that, even though the above prescription may seem natural,
and it does define consistent matching conditions and a unitary evolution,
it should not be given any privileged status. The (compactified) Milne universe
contains a genuine singularity at the origin, and the question of how the
system evolves in the neighborhood of the singularity cannot be in principle
settled through an appeal to a flat Minkowski space (even though there is
nothing wrong with using the covering Minkowski space for constructing
particular evolutionary prescriptions). As we shall see below, more general
rules for singularity crossing can be devised, with a different set
of mode functions and a different vacuum state (which, being an adiabatic
vacuum of infinite order, is no better and no worse than the one
inherited from the covering Minkowski space).

Even though the modefunctions $u_l=\psi_{m,l}^*$ constructed above solve the equations of motion derived from the
action (\ref{actionscalarbis}) at all positive and all negative $t$,
there are no meaningful equations of motion satisfied at $t=0$.
Correspondingly, even though the quantum evolution defined in
terms of the above prescription for the mode functions is unitary
(and essentially inherited from the covering Minkowski space),
this quantum evolution cannot be represented as a solution to
the Schr\"odinger equation for the Hamiltonian derived from (\ref{actionscalarbis}).
In what follows, we shall nevertheless be able to cast this quantum evolution in a Hamiltonian form by appropriately renormalizing the time dependences in the Hamiltonian of the system.

\subsection{Quantum Hamiltonian evolution across the Milne singularity}

In section 2, we constructed a general prescription which allows
to define a Hamiltonian evolution across an isolated singularity in the time
dependence of the Hamiltonian. Since the case of a free scalar field
on the Milne orbifold falls precisely into this category, it will be
instructive to compare the above consideration in terms of the covering
Minkowski space with our general prescription. We shall see that the two
are in fact related, even though it is only in the parametrization of section 2 that the evolution has a manifestly Hamiltonian form at $t=0$.

The Hamiltonian corresponding to the action (\ref{actionscalar}) is
\be
H=\frac1{2|t|}\int dx\,\left(\pi_\phi^2+{\phi^\prime}^2\right)+\frac{m^2|t|}2\int dx\, \phi^2.
\label{hmsingular}
\ee
Following the general guidelines presented in section 2, we shall regulate
the $1/|t|$ time dependence into $f_{1/|t|}(t,\eps)$ of (\ref{f1/t}):
\be
H=\frac1{2}f_{1/|t|}(t,\eps)\int dx\,\left(\pi_\phi^2+{\phi^\prime}^2\right)+\frac{m^2|t|}2\int dx\, \phi^2.
\label{hm}
\ee

Near the origin, where the mass term is negligible, the equations of motion take the form
\be
\ddot\phi-\frac{\dot f_{1/|t|}}{f_{1/|t|}}\,\dot\phi-f_{1/|t|}^2\phi''=0
\ee
or, after Fourier-expanding $\sqrt{2\pi}\phi(x,t)=\sum\phi_l(t)\exp(ilx)$,
\be\label{eomxi}
\ddot\phi-\frac{\dot f_{1/|t|}}{f_{1/|t|}}\,\dot\phi+l^2f_{1/|t|}^2\phi=0.
\ee
The general solution to this equation is
\be
\phi_l=A_l\exp\left[il\int f_{1/|t|}(t,\eps) dt\right]+B_l\exp\left[-il\int f_{1/|t|}(t,\eps) dt\right],
\ee
or
\be
\phi_l=A_l\exp\left[il\left({\rm arcsinh}\frac{t}{\eps}+\frac{2}{\pi}\ln(\mu\eps)\arctan\frac{t}{\eps}\right)\right]+B_l\left[-il\left({\rm arcsinh}\frac{t}{\eps}+\frac{2}{\pi}\ln(\mu\eps)\arctan\frac{t}{\eps}\right)\right].
\ee
With $\eps$ explicitly taken to 0, this becomes
\be
\phi_l=A_l|2\mu t|^{il\,{\rm sign}(t)}+B_l|2\mu t|^{-il\,{\rm sign}(t)}.
\label{0soln}
\ee
To construct the Heisenberg field operator (which contains all information
on quantum dynamics) one should choose any such complex solution and,
after normalizing appropriately, promote it to a mode function,
as in \eq{expansion} (see also appendix A).

The question that will interest us here is how the quantum dynamics
described by the Hamiltonian with our ``minimal subtraction'' is related to the
mode function prescription (\ref{modefunctions}) inherited from the
covering Minkowski space. To this end, we shall define mode
functions $u_l^{(\mu)}$ that solve \eq{eomxi} and coincide with
$u_l$ of \eq{modefunctions} for $t>0$; however, they will
generically differ from $u_l$ for $t<0$. To see the relation between
$u^{(\mu)}_l$ and $u_l$, we construct $u^{(\mu)}_l$ by choosing $A_l$
and $B_l$ in (\ref{0soln}) in such a way that it equals (\ref{hnkl})
for $t>0$ and then compare it, for $t<0$, with (\ref{hankel}).

In order to match (\ref{hnkl}) and (\ref{0soln}) for $t>0$, we impose
\bea
A_l&=&-\frac{1}{2\sqrt{2\pi l\sinh(\pi l)}}\left(\frac{m}{4\mu}\right)^{il}e^{-\frac{\pi l}2-i\varphi_l};\\
B_l&=&\frac{1}{2\sqrt{2\pi l\sinh(\pi l)}}\left(\frac{m}{4\mu}\right)^{-il}e^{\frac{\pi l}2+i\varphi_l}.
\eea
Then, at $t<0$,
\be
u^{(\mu)}_l=\frac{e^{ilx}}{2\sqrt{2\pi l\sinh(\pi l)}}\left(-\left|\frac{8\mu^2t}{m}\right|^{-il}e^{-\frac{\pi l}2-i\varphi_l}+\left|\frac{8\mu^2t}{m}\right|^{il}e^{\frac{\pi l}2+i\varphi_l}\right).
\ee
Comparing this expression with (\ref{hankel}), we conclude that they
are indeed the same if
\be
\mu=\frac{m}4\,\exp\left({-2\varphi_l+\pi\over 2l}\right).
\ee

Note that the fact that $\mu$ depends on the Milne momentum $l$ implies that
the value of the arbitrary parameter introduced by our renormalization procedure
is different for each of the oscillators comprising the field. For that reason,
even though the covering Minkowski space prescription turns out to be the same
as our ``minimal subtraction'' for each of the oscillators, for the entire
field it is not. Phrased in the Hamiltonian language, the covering space
prescription for the Milne singularity transition turns out to be different
from the simplest consistent recipe one could devise, even though it
is related to such simple recipe in a fairly straightforward way.

\section{Conclusions}

We have addressed the issue of how one can define a unitary quantum evolution
in the presence of isolated singularities in the time dependence
of a quantum Hamiltonian. If one demands that the operator structure
of the Hamiltonian should be unaffected by regularization prescriptions (the ``minimal subtraction'' recipe),
one discovers a one-parameter family of distinct quantum evolutions
across the singularity.

For the case of free quantum fields on the Milne orbifold,
the covering Minkowski space considerations
previously brought up in the literature
\cite{Birrell:1982ix, Nekrasov:2002kf, Tolley:2002cv, Berkooz:2002je}
turn out to be closely related to, though distinct from, our ``minimal
subtraction'' proposal.
One explicit advantage of our present approach is that it
makes the evolution across the singularity manifestly Hamiltonian,
which was not the case in the context of the previous discussions.

\section*{Acknowledgments}
We would like to thank D.~Kutasov and N.~Turok for useful discussions. This work was supported in part by the Belgian Federal Science Policy Office through the Interuniversity Attraction Poles IAP V/27 and IAP VI/11, by the European Commission FP6 RTN programme MRTN-CT-2004-005104 and by FWO-Vlaanderen through project G.0428.06.

\appendix

\section{Linear quantum systems}

In this appendix, we shall review the dynamics of linear quantum systems.
This material is very basic and well-known; however, it is usually presented
in relation to a few specific linear systems of physical interest, whereas, for our purposes, it shall be convenient
to summarize here the treatment of a general one-dimensional linear quantum system described by the Hamiltonian
\be
H=\frac{f(t)}{2}P^2+\frac{g(t)}2X^2
\ee
with $f(t)$ and $g(t)$ being arbitrary functions of time.

The equations of motion take the form
\be
\dot P=-g(t)X\qquad \dot X=f(t)P
\ee
or
\be
\ddot X-\frac{\dot f}{f}\,\dot X+fgX=0.
\label{secondorder}
\ee
Should one succeed finding a complex solution $u(t)$ to this equation,
one would be able to write down the most general real solution in
the form
\be
X(t)=Au(t)+A^*u^*(t)
\label{soln}
\ee
with some complex constant $A$. In the quantum case, the solution to the Heisenberg equations of motion will have the exact same form with $A$ and $A^*$
replaced by Hermitean-conjugate operators $a$ and $a^\dagger$:
\be
X_H(t)=au(t)+a^\dagger u^*(t).
\ee

Our solution for the quantum dynamics shall be complete if we establish
the commutation relations for $a$ and $a^\dagger$. Before doing so,
we recall the important notion of Wronskian for a linear
differential equation. For any two solutions $x_1(t)$ and $x_2(t)$
of a second order differential equation, their Wronskian is defined as
\be
W[x_1(t),x_2(t)]=\det\left[\begin{array}{cc}x_1&x_2\vspace{2mm}\\\dot x_1&\dot x_2\end{array}\right].
\ee
It is straightforward to show that, for equation (\ref{secondorder}),
the Wronskian of any two given solutions satisfies
\be
\dot W =\frac{\dot f}f\, W.
\ee
In other words, $W/f$ does not depend on time. This circumstance permits
to define the ``Wronskian norm'' for any complex solution $u(t)$:
\be
\|u\|_{_W}=-i\,\frac{W[u,u^*]}f.
\label{W}
\ee
As we have just demonstrated, the value of this expression does not depend
on the moment of time one chooses to evaluate it. The familiar Klein-Gordon
norm for free quantum fields, which we use in section~3, is a direct
generalization of the Wronskian norm.

The physical relevance of the Wronskian norm becomes apparent
from the consideration of commutators:
\be
-i=[P(t),X(t)]=\frac1{f}\,[\dot X,X]=\frac{\dot uu^*-u\dot u^*}f[a,a^\dagger]=-i\|u\|_{_W}[a,a^\dagger].
\ee
Therefore, to obtain the standard commutation relations for the
creation-annihilation operators, $[a,a^\dagger]=1$, one has to
choose a complex solution $u(t)$ with Wronskian norm $1$.


\end{document}